\documentclass[twocolumn,superscriptaddress,amsmath,amssymb,showpacs,prl]{revtex4-1}

\usepackage{graphicx}
\usepackage{color}
\renewcommand{\phi}{\varphi}

\begin{document}

\title{Phase Diagram and Structure Map of Binary Nanoparticle Superlattices from a Lennard-Jones Model}

\author{Shang Ren}
\affiliation{Department of Physics and Astronomy, Iowa State University, Ames, Iowa 50011, USA}
\affiliation{Ames Laboratory, U.S. Department of Energy, Ames, Iowa 50011, USA}

\author{Yang Sun}
\email{ys3339@columbia.edu}
\affiliation{Department of Physics and Astronomy, Iowa State University, Ames, Iowa 50011, USA}
\affiliation{Ames Laboratory, U.S. Department of Energy, Ames, Iowa 50011, USA}
\affiliation{Department of Applied Physics and Applied Mathematics, Columbia University, New York, NY, 10027, USA}

\author{Feng Zhang}
\email{fzhang@ameslab.gov}
\affiliation{Ames Laboratory, U.S. Department of Energy, Ames, Iowa 50011, USA}

\author{Alex Travesset}
\affiliation{Department of Physics and Astronomy, Iowa State University, Ames, Iowa 50011, USA}
\affiliation{Ames Laboratory, U.S. Department of Energy, Ames, Iowa 50011, USA}
\author{Cai-Zhuang Wang}
\affiliation{Department of Physics and Astronomy, Iowa State University, Ames, Iowa 50011, USA}
\affiliation{Ames Laboratory, U.S. Department of Energy, Ames, Iowa 50011, USA}
\author{Kai-Ming Ho}
\affiliation{Department of Physics and Astronomy, Iowa State University, Ames, Iowa 50011, USA}
\affiliation{Ames Laboratory, U.S. Department of Energy, Ames, Iowa 50011, USA}

\date{June 1, 2020}

\begin{abstract}

A first principle prediction of the binary nanoparticle phase diagram assembled by solvent evaporation has eluded theoretical approaches. In this paper, we show that a binary system interacting through Lennard-Jones (LJ) potential contains all experimental phases in which nanoparticles are effectively described as quasi hard spheres. We report a phase diagram consisting of 53 equilibrium phases, whose stability is quite insensitive to the microscopic details of the potentials, thus giving rise to some type of universality. Furthermore, we show that binary lattices may be understood as consisting of certain particle clusters, \textit{i.e.} motifs, which provide a generalization of the four conventional Frank-Kasper polyhedral units. Our results show that meta-stable phases share the very same motifs as equilibrium phases. We discuss the connection with packing models, phase diagrams with repulsive potentials and the prediction of likely experimental superlattices.

\end{abstract}

\maketitle

Compared with atoms, where size, shape and bonding is completely fixed by the electronic structure, nanocrystals (NCS) offer a degree of tunability as they can be synthesized with any size or shape and may be functionalized with a wide range of ligands\cite{Kovalenko2015}, which determine the bonding and play the same role as electrons in atomic crystals. Just binary NCs systems, for example, form binary nanoparticle superlattices (BNSLs) and quasicrystals of extraordinary complexity\cite{Shevchenko2005, Boles2016}.

Early theoretical treatments described NCs as hard spheres (HS)\cite{Shevchenko2005}, as a clear correlation was found between the maximum of the packing fraction and BNSL stability \cite{Shevchenko2005,Boles2015}. This correlation, however, was rather imperfect, as many experimental systems existed far from the maximum, implying low packing fraction that would likely make those BNSLs unstable. Still, despite its limitations, HS models do provide a natural starting point to describe the equilibrium phases of NC systems: All experimentally reported BNSLs except Li$_3$Bi and AuCu$_3$\cite{Boles2015,Travesset2017a} are thermodynamically stable at the peak of the packing fraction, where each NCs is described as a (quasi)-HS\cite{Travesset2017a}. 

Strict HS models\cite{Eldridge1993,Kummerfeld2008,Filion2009,Filion2009a,Hopkins2012,Damasceno2012,Torquato2018,Cersonsky2018} thus play an important role in the prediction of BNSLs and NC in general. In Ref.~\cite{Travessetpnas2015,HorstTravesset2016,LaCour2019} it was shown that by allowing some compressibility or ``softness'', thus describing NCs as quasi-HS, the thermodynamic stability of the HS binary phases was enhanced and agreement with experiments improved. Based on the softer approximation, the Orbifold Topological Model (OTM)\cite{Travesset2017,Travesset2017a} established the range of validity of the HS approximation, successfully describing all available experimental data as well as subsequent experiments\cite{Coropceanu2019} and simulations\cite{Waltmann2017,Waltmann2018a,Waltmann2018,Zha2018}. These calculations, however, only compared free energies for a set of pre-defined structures, and therefore, the question is how many phases would remain as stable or how many unknown ones would emerge under a general unrestricted structural search. Another important question is that those quasi-HS particles interact through a repulsive potential, thus it is necessary to appeal to the existence of some type of ``universality'' to translate those results into predictions for NC systems. Motivated by these considerations, in this paper we investigate quasi-HS models with attractive interactions. We will therefore use the Genetic Algorithm (GA)\cite{Deaven1995,Ji2010} to perform an open search in systems of Lennard-Jones (LJ) particles with additive interactions. We note that although this paper is motivated by systems in the nanoscale, the results are directly applicable to colloidal systems in the $\mu$-range\cite{Sanders1978,Murray1980} where NCs are well described by quasi-HS throughout.

Another important consideration towards a fully predictive theory for NC structure is the consideration that all experimental BNSLs reported to date can be described as arrangements of a small number of pre-defined particle clusters~\cite{Travesset2017b}, \textit{i.e.} motifs \cite{Sun2016a}, which generalize the four motifs (Z$_{12}$,Z$_{14}$,Z$_{15}$,Z$_{16}$) that describe Frank-Kasper (FK) phases\cite{Frank1958,Frank1959,Kleman1979,Nelson1983}. We will therefore investigate the description of equilibrium and metastable structures as arising from a small subset of motifs as building blocks, not just as a way to construct all possible equilibrium lattices, but also, to identify metastability and glassy or amorphous structures as systems arrested on their way to equilibrium. 

\textbf{Model.} As a minimal model of attractive quasi-HS we consider an
interaction between particles as described by the LJ potential:
\begin{equation}
    U_{LJ} = \left\{
    \begin{aligned}
        &4\epsilon \left(\left(\frac{\sigma}{r}\right)^{12} - \left(\frac{\sigma}{r}\right)^{6}\right) \quad & (r \leq r_{cut}) \\
        &0 \quad & (r > r_{cut})
    \end{aligned}
    \right.
\end{equation}
We consider two types of particles A and B, with the size of A larger than B ($\sigma_{AA} > \sigma_{BB}$). The interaction strength is such that ($\epsilon_{AA}\geq \epsilon_{BB}$), which implements the well documented requirement that the smaller the NCs\cite{Waltmann2017}, the weaker the interaction. All calculations will be performed at $T=0$, and therefore, the parameters $\epsilon_{AA}=1$ and $\sigma_{AA}=1$ are fixed without loss of generality. Then, the system becomes a function of $\gamma = \sigma_{BB} / \sigma_{AA}$, $\sigma_{AB}$, $\epsilon_{AB}$ and $\epsilon_{BB}$. We will further assume that interactions are additive so that the parameters are as follows: 

\begin{widetext}

\onecolumngrid

\begin{align*}
    \sigma_{AA} =  1.0  \qquad & \epsilon_{AA} = 1.0 \\
    \sigma_{AB} = \frac{\sigma_{AA}+\sigma_{BB}}{2} =   \frac{1+\gamma}{2}\qquad & \epsilon_{AB} = \frac{\epsilon_{AA}+\epsilon_{BB}}{2} = \frac{1+\epsilon_{BB}}{2} \\
    \sigma_{BB} =  \gamma \qquad & \epsilon_{BB} = \epsilon_{BB}
\end{align*}

\end{widetext}

\twocolumngrid

In this way, starting from 6 parameters ($\epsilon_{AA}, \epsilon_{BB},\epsilon_{AB},\sigma_{AA},\sigma_{AB},\sigma_{BB}$) the model is reduced to two free parameters ($\epsilon_{BB}$ and $\gamma$). In addition, we introduce the third parameter to control the stoichiometry, which is denoted by $x$. The structures will be presented in the form $A_xB_{1-x}$. In our calculations $\gamma$ is varied from 0.3 to 0.9 and $\epsilon_{BB}$ from 0.1 to 1.0. The step size for both of them is set to be $0.1$. The stoichiometry values are listed in the Table~\ref{tab:my_label}.

The LJ potential has cut-off at a value $r_{cut}$, which was set to be 3.5 times of the radius of the larger particle: $r_{cut} = 3.5 \sigma_{AA}=3.5$. It has been shown that accurate values for thermodynamic quantities are sensitive to the $r_{cut}$\cite{Travesset2014}. One should expect minor corrections on some phase boundaries as a function of the cut-off value, a point that will be elaborated further elsewhere.

\begin{table}[b]
    \caption{Configuration of stoichiometry in GA search}
    \centering
    \begin{tabular}{c|l}
    \hline
    \hline
        x & n(A):n(B)  \\ \hline
        0.1 & 1:9, 2:18 \\ \hline
        0.143 & 1:6, 2:12 \\ \hline
        0.167 & 1:5, 2:10, 3:15 \\ \hline
        0.2 & 1:4, 2:8, 3:12, 4:16 \\ \hline
        0.25 & 1:3, 2:6, 3:9, 4:12, 5:15 \\ \hline
        0.333 & 1:2, 2:4, 3:6, 4:8, 5:10, 6:12 \\ \hline
        0.4 & 2:3, 4:6, 6:9, 8:12 \\ \hline
        0.5 & 1:1, 2:2, 3:3, 4:4, 5:5, 6:6, 7:7, 8:8, 9:9, 10:10 \\ \hline
        0.6 & 3:2. 6:4. 9:6, 12:8 \\ \hline
        0.667 & 2:1, 4:2, 6:3, 8:4, 10:5, 12:6 \\ \hline
        0.75 & 3:1, 6:2, 9:3, 12:4, 15:5 \\ \hline
        0.8 & 4:1, 8:2, 12:3, 16:4 \\ \hline
        0.833 & 5:1, 10:2, 15:3 \\ \hline
        0.857 & 6:1, 12:2 \\ \hline
        0.9 & 9:1, 18:2 \\ \hline
        \hline
    \end{tabular}

    \label{tab:my_label}
\end{table}

\section{Results and Discussion}

We first illustrate the method in some detail for the case $\epsilon_{BB}=1.0$, and then present the general results. We also proceed to rigorously characterize the motifs and identify them in the lattice structures.

In order to name the different phases we searched the Material Project Database\cite{Jain2013} to find a prototype isostructural phase and name the GA calculated lattice accordingly. If no match is found, then we name the phase according to the following convention:
\begin{equation}\label{Eq:naming_convention}
{\mbox{A}_m \mbox{B}_n}^{\mbox{space group}}_{\mbox{identifier}} .
\end{equation}
Here $m$ and $n$ are the number of A and B particles within the unit cell. The space group is determined using the FINDSYM package\cite{Stokes2005}, with the tolerance for lattice and atomic positions set to $0.05$. The identifier is necessary as multiple phases with the same stoichiometry and space group, differing only in Wyckoff number and positions, are found.

\begin{figure}[t]
\includegraphics[width=0.48\textwidth]{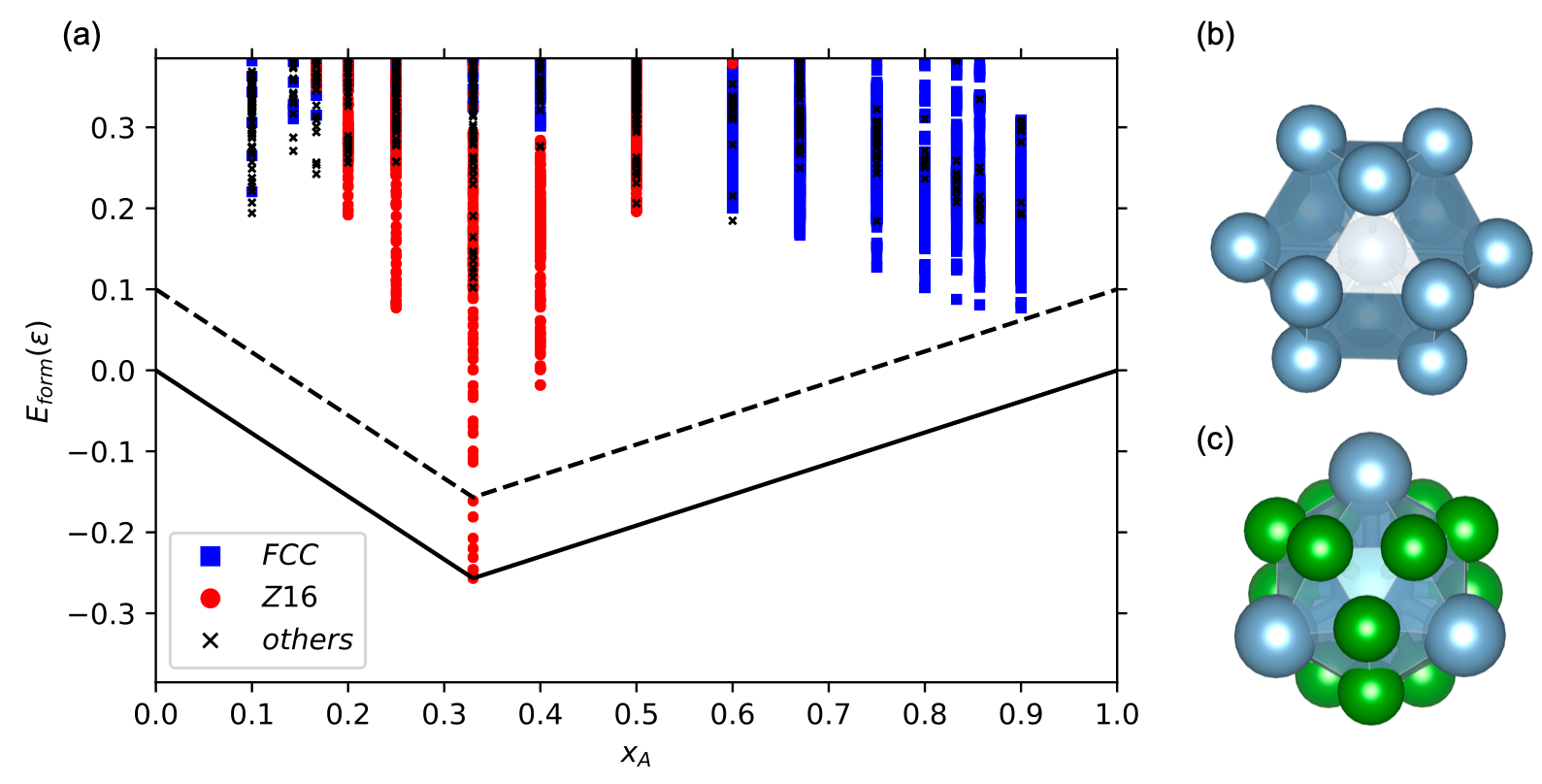}
\caption{Structures searched by GA in $\epsilon_{BB}=1.0$ and $\gamma = 0.8$. (a) Formation energies ($E_{form}$) of structures searched by GA as a function of stoichiometry ($x$). Each point corresponds to a structure. The color of points are assigned by the type of motifs in the corresponding structure. The black solid line is the convex hull of the system, while the black dash line is the threshold for metastable structures. (b) Structure of the FCC motif (c) Structure of the MgZn$_2$ motif, which is Frank-Kasper Z$_{16}$.}\label{Fig:epsilon_1}
\end{figure}

\textbf{The case $\boldsymbol{\epsilon_{BB}=1.0}$.} Here we consider $\epsilon_{BB} = \epsilon_{AA} = 1.0$, while $0.3 \le \gamma \le 0.9$. We first compute the energy of the ground state for the pure A and B states, which previous calculations\cite{Stillinger2001,Travesset2014} have shown to be the hcp phase. Here, however, because of the finite cut-off of LJ potentials, the fcc phase has lower energy. The identification of equilibrium phases proceeds by comparing their energy against phase separation into pure $A$ and $B$. Then, out this list of putative binary phases that are stable against phase separation, the energies are compared to establish the resulting true phase diagram equilibrium. This is how the phase diagram Fig.~\ref{Fig:epsilon_1} is built, where there is only one stable BNSL, the MgZn$_2$ Frank-Kasper phase at $\gamma=0.8$. We should note that maximum of the packing fraction for this phase occurs for $\gamma_c=\sqrt{2/3}=0.8165$\cite{Travesset2017a}, which is very close.

Since it is common that structures that are metastable at 0 K can be observed in experiments at finite temperatures, we also considered metastable phases defined to be those within $0.1\epsilon$/particle in energy above the convex hull. As shown in Fig.~\ref{Fig:epsilon_1}, there are a number of metastable phases at $x=0.333$, which are minor variations of MgZn$_2$ as we analyze further below in the context of motifs.

\begin{figure}[b]
    \includegraphics[width=1.0\textwidth]{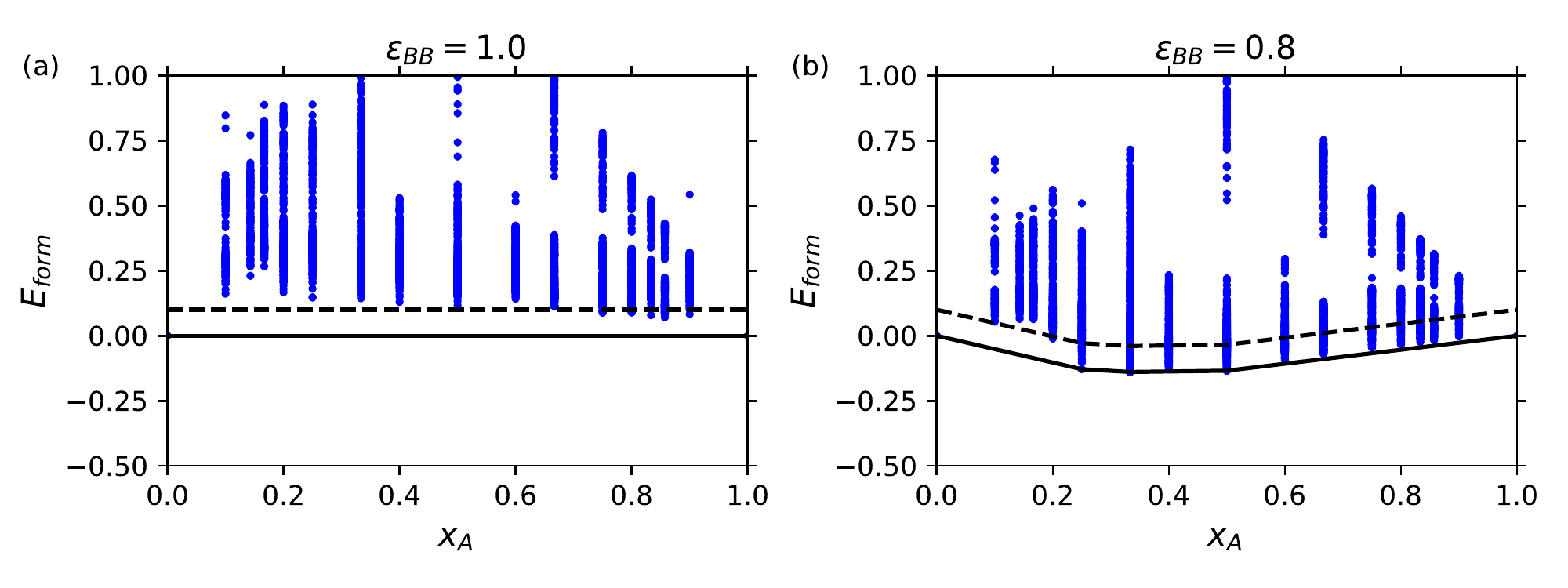}
    \caption{Two examples of GA results for $\gamma = 0.6$. In each figure, the solid line is the convex hull, while the dashed line is the threshold for metastable structures, see the discussion above. (a) Structures searched by GA as a function of x when $\epsilon_{BB} = 1.0$: There are no stable binary structures between x=0 and x=1. (b) Structures searched by GA as a function of x for $\epsilon_{BB}=0.8$. There are three stable structures which appear at $x = 0.25$ (TiCu$_3$),  $x = 0.333 $ (AlB$_2$) and $x = 0.5$ (CrB).}\label{Fig:epsilon_gen_energy}
\end{figure}

\textbf{General $\epsilon_{BB}$.} On physical grounds, it is expected that the smaller the particle the weaker the interaction, hence we consider $\epsilon_{BB} \le 1$. In Fig.~\ref{Fig:epsilon_gen_energy}, we provide a typical calculation for fixed $\gamma=0.6$ as a function of both $\epsilon_{BB}$ and $x$. As expected, see  Fig.~\ref{Fig:epsilon_1}, the phase diagram is trivial for $\epsilon_{BB}=1$. However, three phases TiCu$_3$, AlB$_2$ and CrB at $x=0.25, 0.333, 0.5$ are found for $\epsilon_{BB}= 0.8$.

\begin{figure}[]
\includegraphics[width=1.0\textwidth]{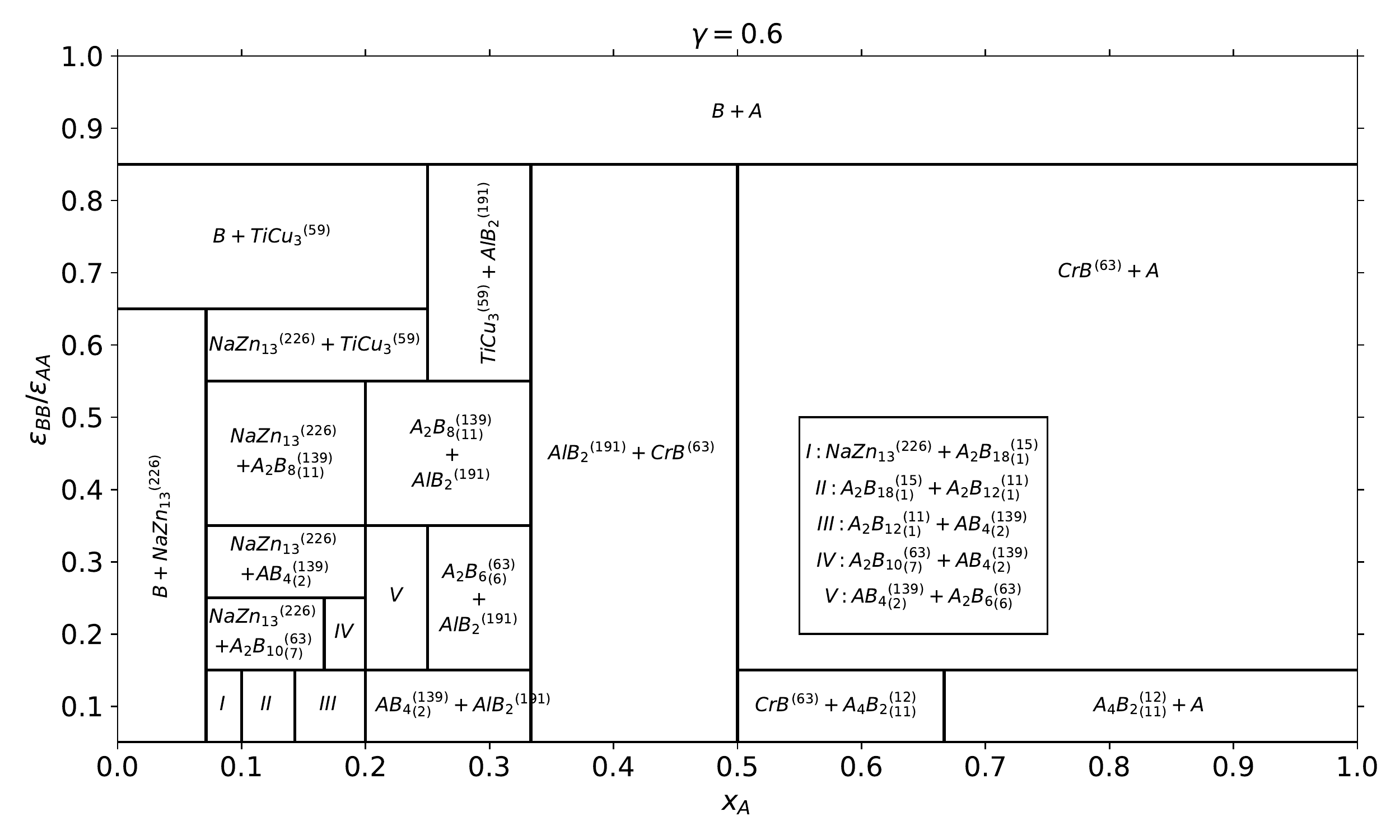}
\caption{Phase diagram in $x$ and $\epsilon_{BB}/\epsilon_{AA}$ for $\gamma = 0.6$.}\label{Fig:epsilon_gen}
\end{figure}

By repeating the calculations shown in Fig.~\ref{Fig:epsilon_gen_energy} for the other values of $\epsilon_{BB}$ at a fixed $\gamma=0.6$ (see Table~\ref{tab:my_label}), we constructed the phase diagram shown in Fig.~\ref{Fig:epsilon_gen}. In Fig.~\ref{Fig:epsilon_gen} we note the appearance of seven additional phases for $\epsilon_{BB}< 0.6$ that could not be matched to any prototype: Detailed description for these and all other equilibrium phases are collected in Supporting Information Table S1. A database for all the structures is included in Supporting Information.

Similarly, the phase diagrams for all other values of $\gamma$ are also presented in Supporting Information Fig. S2. Common to all these phase diagrams is the appearance of many diffusionless (martensitic), usually incongruent transformations, as a function of the  energy parameter $\epsilon_{BB}/\epsilon_{AA}$. In Supporting Information Fig. S3, we have also included phase diagrams for all values of $\epsilon_{BB}/\epsilon_{AA}$ in $x$ and $\gamma$.

\textbf{Motifs.} We define motifs as the polyhedron consisted of a center particle and its first-shell neighbors. The motifs are generated according to the analysis of bond length table from neighboring particles to the center (see details in Supporting Information Fig. S6). In this study, we only include motifs with the larger A-particles as the center. We will name motifs according to
\begin{equation}\label{Eq:motif_name}
    \mbox{Motif}-\mbox{CN}-\mbox{Identifier} \ ,
\end{equation}
where CN is the Coordination (the number of particles) and identifier discriminates among motifs with the same coordination number.

\onecolumngrid

\clearpage

\twocolumngrid

We identified 187 equilibrium and 102,822 metastable structures. Out of the 187 equilibrium structures, we removed redundancies by a cluster alignment algorithm\cite{Fang2010,Sun2016a}leading to only 53 equilibrium structures. Out of these 53 structures we identified 42 motifs, which are listed in the order of increasing CN in Supporting Information Fig. S4. 416,391 motifs can be found in the 102,822 metastable structures. Among them, a vast majority (312,891) of the motifs of the metastable structures also exist in the equilibrium phases. In Tab~\ref{tab:motif}, we list the name, CN and the percentage fraction of the ten most frequent motifs present in meta-stable structures. Note that these ten already account for more than 95\% of the 312,891 motifs. The details about how to identify the motif from a crystal and how to identify if a crystal has the motif inside have been included in the Supporting Information.

\begin{table}[h]
    \centering
    \caption{Ten most frequent Motifs in metastable structures}
    \begin{tabular}{c|c|c}
    \hline
    \hline
        Motif & CN & Frequency \\ \hline
        FCC & 12 & 31.4\% \\ \hline
        HCP & 12 & 18.9\% \\ \hline
        Octahedron (Motif-6-4) & 6 & 9.8\% \\ \hline
        Half Hexagonal Prism 1 (Motif-6-2) & 6 & 9.1\% \\ \hline
        Triangular Prism (Motif-6-3) & 6 & 6.4\% \\ \hline
        Half Hexagonal Prism 2 (Motif-6-1) & 6 & 5.3\% \\ \hline
        BCC & 8 & 5.0\% \\ \hline
        Hexagonal Prism (Motif-12-3) & 12 & 4.8\% \\ \hline
        Half Truncated Cube (Motif-12-1) & 12 & 2.3\% \\ \hline
        MoB (Motif-13-1) & 13 & 2.2\% \\ \hline
        Total & & 95\% \\ \hline
        \hline
    \end{tabular}
    \label{tab:motif}
\end{table}

As an illustrative example, we consider the case of $\epsilon_{BB}=1$ and $\gamma=0.8$, where in Fig.~\ref{Fig:epsilon_1} we have shown the two relevant motifs are the FCC and the Frank-Kasper Z$_{16}$. 
By coloring each structure according to the motif, we can confirm that the metastable phases (all in red) have motifs which are small variations of the Frank-Kasper Z$_{16}$ and that the vast majority of the structures found in other searches have motifs which are variations of either FCC or Frank-Kasper Z$_{16}$.

\onecolumngrid

\begin{figure}[]
\includegraphics[width=0.7\textwidth]{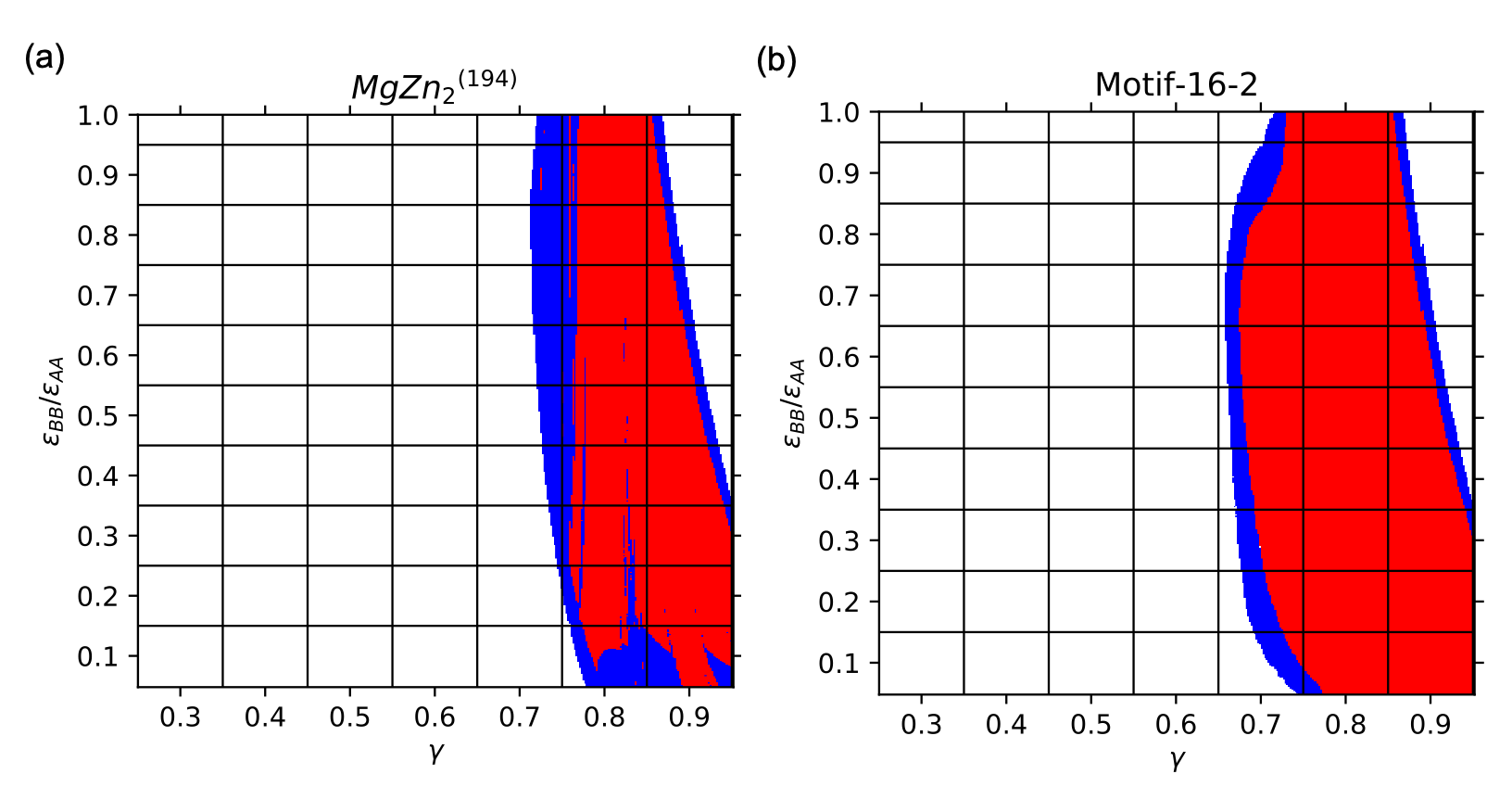}
\caption{(a) Map of MgZn$_2$ in $\gamma$ and $\epsilon_{BB}$. The red regime indicates that the structure of MgZn$_2$ is thermodynamically stable while in blue regime it is metastable. (b) Map for the Z$_{16}$ motif with red stable, blue metastable. The red regime is where the stable structure has Z$_{16}$ motif inside. Note that the motif has a wider range of both stability and metastability, as it also appears in other Laves phase, such as the MgCu$_2$ and MgNi$_2$.}\label{Fig:mgzn2_motif_map}
\end{figure}

\twocolumngrid

\onecolumngrid

\begin{figure}[]
    \includegraphics[width=0.8\textwidth]{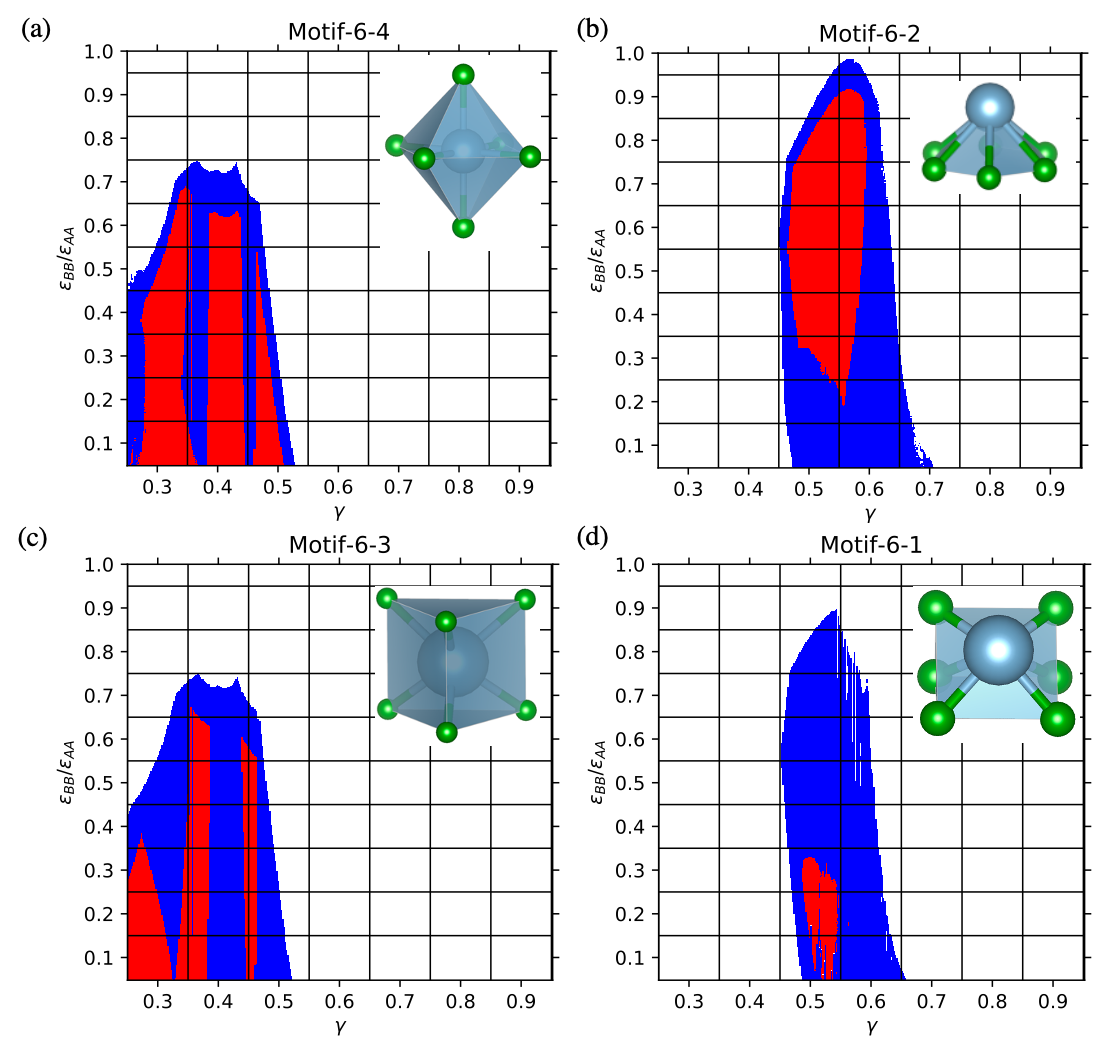}
    \caption{Map for first four frequent motifs in $\gamma$ and $\epsilon_{BB}$ excludes the general motif FCC and HCP. (a) Octahedron (b) Half hexagonal prism 1. (c) Triangular Prism (d) Half hexagonal prism 2. Red indicates stable structures and blue indicates metastable.
    }\label{Fig:req_motif}
\end{figure}

\twocolumngrid

In Fig.~\ref{Fig:mgzn2_motif_map} we show the domain of stability and metastability for the MgZn$_2$ phase and the Z$_{16}$ motif. The GA searches were performed on a mesh of $\gamma$ and $\epsilon_{BB}$ with an increment of 0.1. Here, to improve the resolution of the stability range, we examined the stability of all GA-found structures and motifs on a finer mesh in the $\gamma$-$\epsilon_{BB}$ plane with an increment of 0.02. Rather interestingly, the stability range of the Z$_{16}$ motif is larger than that of the MgZn$_2$ phase, indicating this motif is not unique to MgZn$_2$, but shared by other Laves and Frank-Kasper phases. Similar plots for the four more frequent motifs are shown in Fig.~\ref{Fig:req_motif}.

\begin{table}[h]
    \centering
    \caption{Comparison between packing phases\cite{Hopkins2012} and our study for $ 0.3 \le \gamma  \le 1$. (SG=Space group), The $\ast$ indicates there are small distortions in the LJ phase, compared with the packing phase. Motifs in the LJ column indicates that they are not stable in the GA result, but they have the motif inside in the corresponding $\gamma$ regime.}
    \begin{tabular}{c c c c | c c}
    \hline
    \hline
     Phase   & $\gamma$-range        &  Ref  &  SG &  LJ & Distortion \\ \hline
     A$_3$B  & $[0.618, 0.660]$ & \cite{OToole2011} &  59  & TiCu$_3$  &  \\
     AlB$_2$  & $[0.528, 0.620]$  &       & 191 & AlB$_2$ & \\
     AuTe$_2$ & $[0.488, 0.528]$ & \cite{Filion2009} & 12 & Motif-6-2 &   \\
     (2-2)$^{\ast}$ & $[0.480, 0.497]$ &\cite{Marshall2010} & 11 & Motif-6-1 & \\
     (4-2) & $[0.488, 0.483]$ &\cite{Hopkins2012} & 191 & Motif-12-3 & \\
     (5-2) & $[0.480, 0.483]$ &\cite{Hopkins2012} & 44 & & \\
     (7-3) & $[0.468, 0.480]$ &\cite{Hopkins2012} & 71 & Motif-12-3 & \\
     HgBr$_2$ & $[0.443, 0.468]$ &\cite{Filion2009} & 36 & Motif-6-4 & \\
     (6-6) & $[0.414, 0.457]$ &\cite{Hopkins2012} & 11 & Motif-6-4 &  \\
     XY & $[0.275, 0.414]$ &\cite{Hopkins2012} &  & & \\
     (6,1)$_4$ & $[0.352, 0.321]$ &\cite{Hopkins2012} & 69 & ${\mbox{A}_2\mbox{B}_{12}}^{(139)}_{(1)}$ & $\ast$ \\
     (6,1)$_6$ & $[0.321, 0.304]$ &\cite{Hopkins2012} & 139 & ${\mbox{A}_2\mbox{B}_{12}}^{(139)}_{(1)}$ & $\ast$ \\
     (6,1)$_8$ & $[0.302, 0.292]$ &\cite{Hopkins2012} & 139 & ${\mbox{A}_2\mbox{B}_{12}}^{(139)}_{(1)}$ & $\ast$ \\
     \hline\hline
     \\
    \end{tabular}

    \label{tab:pg_lj_comp}
\end{table}

Quite generally, the motifs are far more sensitive to $\gamma$ than they are to $\epsilon_{BB}/\epsilon_{AA}$, confirming that the particle size is more important than the actual intensity of the interactions. It is consistent with all calculations that stable structures with the same values of $\gamma$ tend to share motifs. As found for MgZn$_2$ and Z$_{16}$, the regions for stability and metastability is wider than the corresponding structures, thus indicating that motifs define very general families of structures, like Laves phases. A classification of motifs by Renormalized Angle Sequences (RAS)\cite{Lv2017,Lv2018} has been included in Supporting Information. 

This study has identified 53 equilibrium lattices and 42 motifs (with the larger particle A as reference). We now discuss the relevance of these results for packing models\cite{Filion2009,Hopkins2012}, their connection to the motifs reported in Quasi Frank-Kasper phases\cite{Travesset2017} and their implications for binary superlattices.

\textbf{Packing Phase Diagram.} We consider the study of Hopkins \textit{et al.}\cite{Hopkins2012} as the reference phase diagram for packing problems, although it only includes unit cells containing up to 12 particles. Consistently with this study we concentrate on the range $0.3\le \gamma \le 1$, also because for smaller $\gamma$ there are many phases with narrow stability ranges that are less relevant in actual experimental systems.

From Table~\ref{tab:pg_lj_comp}, the packing of binary phase diagram contains 13 phases for the $0.3 \le \gamma \le 1$ range. For large $\gamma > 0.528$ only two phases exist; AlB$_2$ and A$_3$B, which are both found in binary LJ systems  (if allowing for small differences in A$_3$B). For $0.488 < \gamma < 0.528$, however, the AuTe$_2$ phase is reported; We did not find such phase, but we do report the Motif-6-2 as stable for the same range of $\gamma$, see Supporting Information, which is present in the equilibrium phases at $\gamma=0.5$ ${{\mbox A}_4{\mbox B}_6}^{(166)}_{(9)}$, BaCu and TePt. Some other phases, which are reported as packing phases \cite{Hopkins2012} but not stable in the GA search, are also identified to have the motif in the corresponding $\gamma$ regime. This indicates that these packing phases may be meta-stable in our calculation. For smaller $\gamma$, there is also overlap if allowing for small distortions. 

Other phases that have large packing fractions, such as CrB and S74e/h(KHg$_2$ in our notation)\cite{Filion2009}, that are metastable in the packing phase diagram become equilibrium, thus showing that the LJ system augments the number of stable phases as compared with packing models.

\textbf{Motifs and Quasi Frank Kasper Phases.} In Ref.~\cite{Travesset2017} it was shown that all experimental BNSLs could be described as disclinations of the $\{3,3,5\}$ polytope, thus generalizing well known four Frank-Kasper motifs Z$_{12}$,Z$_{14}$,Z$_{15}$, Z$_{16}$\cite{Frank1958,Frank1959} to include other motifs. 

\onecolumngrid

\begin{table}[]
    \centering
    \caption{Motifs in Quasi Frank Kasper phases\cite{Travesset2017} compared to the ones described in this work.} 
    \begin{tabular}{c | c c c c c c}
    \hline\hline
    QFK\cite{Travesset2017}  & $\mbox{Z}_6$ & $\mbox{Z}_{12}^{\prime\prime}$ & $\mbox{Z}_{14}^{\prime}$  & $\mbox{Z}_{16}$ & $\mbox{Z}_{18}^{\prime\prime}$ & $\mbox{Z}_{24}$ \\
    \hline
    This work & Motif-6-4 & Motif-12-2 & Motif-14-1 & Motif-16-2 & Motif-18-3 & Motif-24-1 or  \\
              &        &         &         &         & & Motif-24-3 \\
    \hline
    \hline

    \end{tabular}
    \label{tab:qfk_lj_comp}
\end{table}

\twocolumngrid

In Table~\ref{tab:qfk_lj_comp} we show the equivalence between Quasi Frank Kasper motifs and the ones obtained in this work, which only include those with the A-particle as reference. It should be pointed that the motifs are not completely the same, as in Ref.~\cite{Travesset2017} the motifs were defined by the Voronoi cell and its corresponding neighbors, which is a slightly different definition than the one used in this paper. 

\textbf{Experimental Results.} The list of experimentally reported BNSLs is taken from Ref.~\cite{Travesset2017a}, where we have excluded two dimensional superlattices and those where nanocrystals cannot be approximated as spherical, see Ref.~\cite{Boles2016}. The comparison between the results obtained in this paper and experimental BNSLs is provided in Table~\ref{tab:exp_lj_comp}.

\begin{table}[]
    \centering
    \caption{Experimentally determined structures. NA: Phase not available in this study. NF: Phase not found in this study. The DDQC/AT is a quasicrystal phase. The bccAB$_6$ phase is also known as C$_{60}$K$_6$ and is denoted as ${\mbox{A}\mbox{B}_6}^{(229)}_{(1)}$ in this paper.} 
    \begin{tabular}{c c | c c c}
    \hline \hline
    \multicolumn{2}{c|}{Experiment} &  \multicolumn{3}{c}{Binary LJ}\\ \hline
    BNSL & $\gamma$-range &     & $\gamma$-range & $\varepsilon_{BB}$-range  \\
    NaCl &  $[0.41,0.60]$  &    &  $[0.2,0.5]$    & $[0.1,0.8]$     \\
    CsCl &  $[0.71,0.90]$  &    &   NF   &      \\
    AuCu &  $[0.58,0.71]$  &    &    NF  &      \\
    DDQC/AT&  $[0.41,0.43]$  &    &  NA   &      \\
    AlB$_2$ &  $[0.45,0.70]$  &    & $[0.4,0.7]$    & $[0.1,0.9]$     \\
    MgZn$_2$ & $[0.60,0.81]$  &    & $[0.7,1.0]$     & $[0.1,1.0]$  \\
    AuCu$_3$ & $[0.40,0.60]$  &   &   NF   &      \\
    Li$_3$Bi & $[0.53,0.56]$  &   &   NF   &      \\
    Fe$_4$C  & $[0.55,0.65]$  &    &   NF   &     \\
    CaCu$_5$ & $[0.60,0.80]$  &    &  $[0.6,0.8]$    & $[0.1,0.9]$    \\
    CaB$_6$  & $[0.43,0.47]$  &    & $[0.3,0.5]$    & $[0.1,0.8]$    \\
    bccAB$_6$  & $[0.45,0.50]$  &    & $[0.4,0.6]$      & $[0.1,0.5]$     \\
    cubAB$_{13}$  & $[0.55,0.60]$  &   &  NF    &     \\
    NaZn$_{13}$  & $[0.47,0.70]$  &    & $[0.6]$  & $[0.1,0.6]$     \\
    \hline \hline
    \end{tabular}
    \label{tab:exp_lj_comp}
\end{table}

Seven of the experimentally reported BNSLs, namely NaCl, AlB$_2$, MgZn$_2$, CaCu$_5$, CaB$_6$, bccAB$_6$ and NaZn$_{13}$ are found as equilibrium phases in the LJ system essentially for the same range of $\gamma$. The fact that in our results the stability is roughly independent of $\varepsilon_{BB}$ in certain regions provides some support for the idea that microscopic details of the potential are unimportant in this region (``universality''). Further making this point is that the same phases are stable for soft repulsive potentials in the same $\gamma$-range \cite{Travessetpnas2015,HorstTravesset2016,LaCour2019}. 

\begin{figure}[]
    \includegraphics[width=0.48\textwidth]{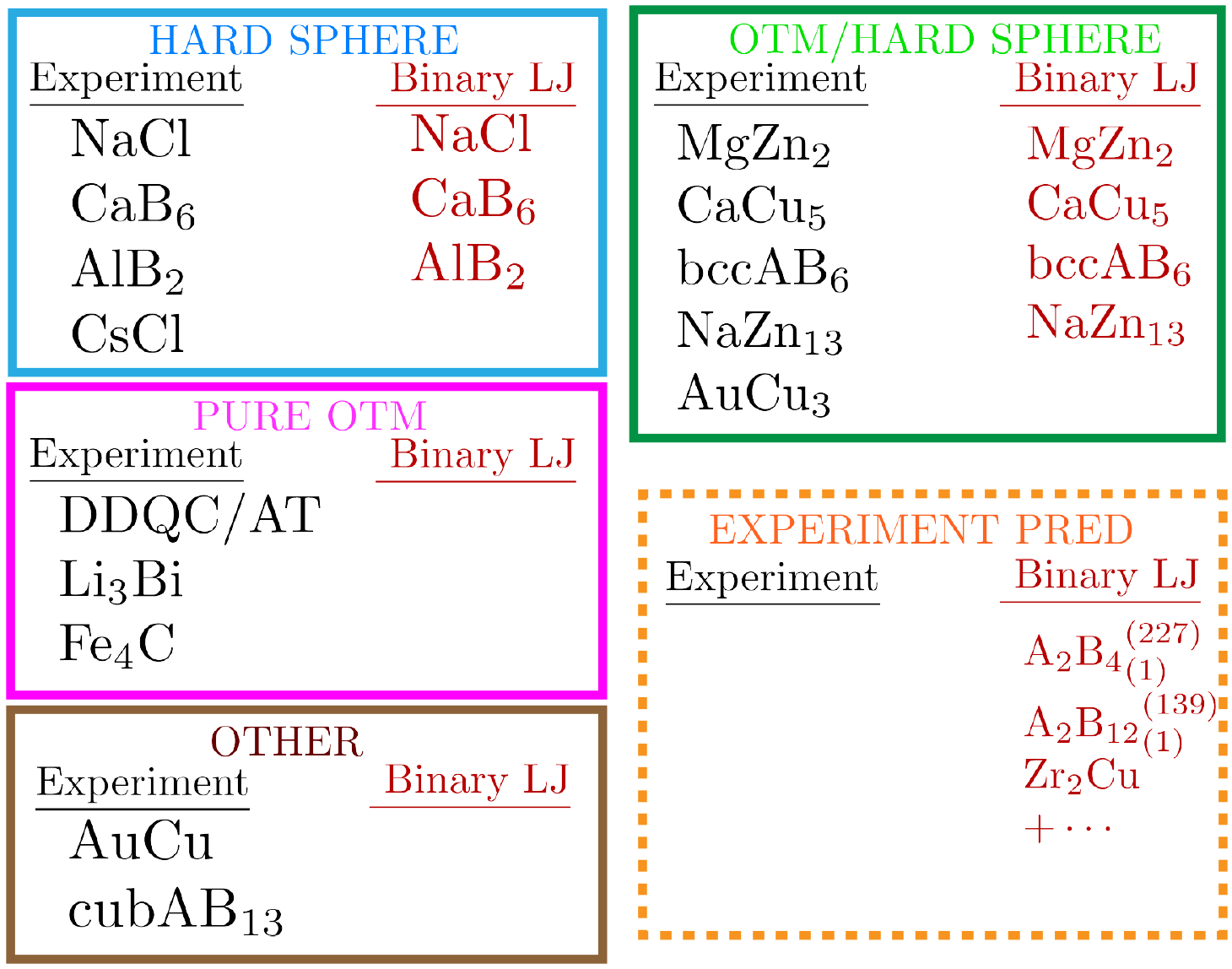}
    \caption{Summary of the main results of the paper: The experimental phases are classified according to: Hard sphere, OTM/hard sphere (exist when NCs are modeled as hard spheres but are stabilized by vortices)\cite{Travesset2017}, pure OTM(only stable with vortices), and other (observed in special cases, such as ligand detachment\cite{Boles2019}). See also Table~\ref{tab:exp_lj_comp}. Consistent with the LJ assumptions, only the hard sphere phases are found in our work. The Experiment Pred includes those strong candidates to be found experimentally, as discussed below.}\label{Fig:summary}
\end{figure}

We now analyze the phases reported in experiments that are not equilibrium in our study. One of them is beyond the scope of our calculation; DDQC/AT, which is a quasicrystal. The Li$_3$Bi and also the AuCu$_3$ are stabilized by large deformations of the ligands, i.e. vortices\cite{Travesset2017}, and therefore are not possible to obtain from a quasi HS approximation. The Fe$_4$C phase was observed in 2006\cite{Shevchenko2006}, and since then, it has not been reported in any further study, which may suggest is metastable, and furthermore, it can only be stabilized by vortices\cite{Travesset2017a}. The CsCl phase has a very narrow range of stability around $\gamma_c = \sqrt{3}-1=0.732$\cite{Travesset2017a}, which is likely missed by the discretization of $\gamma$ values in our study.
Finally, AuCu occurs when there is ligand loss\cite{Travesset2017,Boles2019} and is stabilized through a different mechanism involving the non-spherical shape of the nanocrystal. We therefore conclude that the binary LJ model successfully predicts those experimentally reported phases that can be described as quasi-hard spheres. This is in contrast to packing models, where MgZn$_{2}$ or CaCu$_5$ phases, widely reported in experiments are not equilibrium phases (maximum of the packing fractions). See Fig.~\ref{Fig:summary} for a visual summary of this discussion.

\section{Conclusions}

By the use of Genetic Algorithm (GA), we have been able to predict stable structures under different sizes of particles and strengths of interaction ($\gamma \in$  [0.3 to 0.9], $\epsilon_{BB} \in$ [0.1 to 1.0]). We report 53 stable phases, which cover a significant part of currently reported structures.  Besides that, we also predict 35 stable structures which are not in Material Project database. We find that the type of stable structures strongly depends on $\gamma$, but weakly on $\epsilon_{BB} < 1$, providing evidence that the stability of the lattices has a weak dependence on the potential details (universality). By comparing our results with other theoretical and experimental works, it is shown that regardless of potential details, the same $\gamma$ regime has the same stable structure, which reinforce that the stable structure has a weak dependence on the potential details.

There are two aspects about the limitations of the hard sphere description: The first is that it does \textit{not} provide a free energy: the observed phases are not the ones with maximum packing fraction\cite{Hopkins2012}, but rather, ones where the packing fractions is maximum for the particular structure. This is where the Binary LJ becomes important: the stable phases are the ones that minimize the free energy (modeled as the LJ potential). The second limitation is that it does not model large deformations of the ligand shell: these cases go beyond the LJ model and is evident from Fig.~\ref{Fig:summary}, showing that these phases are absent.

The crystalline motifs are employed to describe the large amount of metastable structures. We find that metastable structures mostly can be described from the motifs present in equilibrium structures, thus suggesting the possibility of building superlattices by patching all motifs that can tile the 3D space, as similarly done in the more restricted case of Frank-Kasper phases\cite{DutourSikiric2010}. It also raises the possibility of motifs being present within the liquid\cite{Damasceno2012} as a way to anticipate the emergent crystalline structure.

Comparing with available experimental results, see Table~\ref{tab:exp_lj_comp} and Fig.~\ref{Fig:summary}, the binary LJ model captures all the equilibrium phases where nanocrystals can be faithfully described as quasi hard spheres: NaCl, AlB$_2$, MgZn$_2$, CaCu$_5$, CaB$_6$, bccAB$_6$ and NaZn$_{13}$. The other phases reported in experiments either require the presence of vortices, as predicted by the OTM\cite{Travesset2017,Travesset2017a}, or are stable over a very narrow range of $\gamma$ values, likely missed by the necessary discrete number considered in our study.

Packing phase diagram models reported 14 equilibrium phases in the interval $\gamma \in [0.3,1)$, see Table~\ref{tab:pg_lj_comp}, while our study reports 53, thus showing that binary LJ have a more complex phase diagram. Rather interestingly, phases such as MgZn$_2$ or CaCu$_5$, which are very common in experiments, are absent in the packing phase diagram; Although very useful in identifying at which $\gamma$ values a phase is likely to appear, packing models give very poor predictions on which, among all possible phases, will actually be observed.

The two guiding principles for stability of BNSLs in experiments are high packing fraction (or low Lennard-Jones Energy) and tendency towards icosahedral order, as reflected in the motifs\cite{Travesset2017a,Coropceanu2019}. Therefore, we expect that those equilibrium Lennard-Jones phases with Quasi Frank-Kasper motifs, 
for example, the BNSLs ${\mbox{A}_{2}\mbox{B}_{4}}^{(227)}_{(1)}$ and
${\mbox{A}_{2}\mbox{B}_{12}}^{(139)}_{(1)}$
(Motif-16-2), or $\mbox{Zr}_{2}\mbox{Cu}^{(139)}_{(1)}$ (Motif-14-1), will be excellent candidates to search for BNSLs,  see Fig.~\ref{Fig:summary}. Definitely, these ideas will be developed further in the near future, where the 53 stable lattices will be studied with more realistic nanocrystal models described at the atomic level. 

In this work we focused on spherically symmetric potentials with additive interactions, as described by relations like
\begin{equation}
\epsilon_{AB}=\frac{1}{2}(\epsilon_{AA}+\epsilon_{BB}) \ .
\end{equation}
It is of interest to consider more general models, where these restrictions are lifted. This, however, will be the subject of another study.

\section{Methods}
The crystal structure searches with GA were only constrained by stoichiometry, without any assumption on the Bravais lattice type, symmetry, atom basis or unit cell dimensions (up to a maximum of particles per unit cell). During the GA search, energy was used as the only criteria for optimizing the candidate pool. At each GA generation, 64 structures are generated from the parent structure pool \textit{via} the mating procedure described in Ref.~\cite{Deaven1995,Oganov2006,Ji2010}. The mating process was based on real-space “cut-and-paste” operations that was first introduced to optimize cluster structures~\cite{Deaven1995}. This process was extended to predict low-energy crystal structures by Oganov~\cite{Oganov2006} and reviewed in Ref.~\cite{Ji2010}. Here, we follow the same procedure that was described in detail in Ref.~\cite{Ji2010} and was implemented in the Adaptive Genetic Algorithm (AGA) software.

With a given set of LJ parameters, we performed three GA searches independently, with each GA search running for 1000 generations. The maximum number of particles per unit cell used in each search was 20, and thus, phases with large unit cells, the most relevant being NaZn$_{13}$, could not be included. Therefore, we include NaZn$_{13}$ into our calculation manually. All energy calculations and structure minimizations were performed by the LAMMPS code \cite{Plimpton1995} with some cross checks using HOOMD-Blue\cite{AndersonMe2008a} with FIRE minimization\cite{Bitzek2006}. The database of binary lattices in HOODLT\cite{Travesset2014} was also used.

\section{Supporting Information}
Supporting information contains: 
List and maps of structures searched by genetic algorithm;
phase diagrams of equilibrium structures; equilibrium motif database; maps of motifs; algorithms for motif identification and renormalized angle sequence

\section{acknowledgement}

A.T acknowledges discussions with I. Coropceanu and D. Talapin. We also thank Prof. Torquato for facilitating the data of his group packing studies. Work at Ames Laboratory was supported by the US Department of Energy, Basic Energy Sciences, Materials Science and Engineering Division, under Contract No. DE-AC02-07CH11358, including a grant of computer time at the National Energy Research Supercomputing Center (NERSC) in Berkeley, CA. The Laboratory Directed Research and Development (LDRD) program of Ames Laboratory supported the use of GPU-accelerated computing. Y. S. was partially supported by National Science Foundation award EAR-1918134 and EAR-1918126.

\bibliographystyle{apsrev4-1}
\bibliography{ref}

\end{document}